\documentclass[aps,prb,superscriptaddress,amsfonts,amsmath,amssymb,twocolumn,showpacs,floatfix]{revtex4-1}
\usepackage{url}
\usepackage{bm}
\usepackage{graphicx}
\usepackage{amsmath}
\usepackage{amstext}
\usepackage{amssymb}
\usepackage{amsfonts}
\usepackage{amsbsy}
\usepackage{verbatim}
\usepackage{color}
\usepackage[colorlinks=true, urlcolor=blue, linkcolor=blue, citecolor=blue, pdftex]{hyperref}
\usepackage{multirow}

\begin{document}

\title{Valence-bond crystal in the extended Kagome spin-$\frac{1}{2}$ quantum 
Heisenberg antiferromagnet: A variational Monte Carlo approach}

\author{\href{http://www.lpt.ups-tlse.fr/spip.php?article38&lang=en}{Yasir Iqbal}}
\email[]{yasir.iqbal@irsamc.ups-tlse.fr}
\affiliation{Laboratoire de Physique Th\'eorique UMR-5152, CNRS and 
Universit\'e de Toulouse, F-31062 France}
\author{\href{http://www.democritos.it/curri/federico.becca.php}{Federico Becca}}
\email[]{becca@sissa.it}
\affiliation{Democritos National Simulation Center, Istituto
Officina dei Materiali del CNR and Scuola Internazionale Superiore di Studi 
Avanzati (SISSA), Via Bonomea 265, I-34136 Trieste, Italy}
\author{\href{http://www.lpt.ups-tlse.fr/spip.php?article32}{Didier Poilblanc}}
\email[]{didier.poilblanc@irsamc.ups-tlse.fr}
\affiliation{Laboratoire de Physique Th\'eorique {\rm UMR-5152}, {\rm CNRS} and 
Universit\'e de Toulouse, {\rm F-31062 France}}

\date{\today}

\begin{abstract}
The highly-frustrated spin-$\frac{1}{2}$ quantum Heisenberg model with both nearest
($J_1$) and next-nearest ($J_2$) neighbor exchange interactions is revisited 
by using an extended variational space of projected wave functions that are 
optimized with state-of-the-art methods. Competition between modulated
valence-bond crystals (VBCs) proposed in the literature and the Dirac spin 
liquid (DSL) is investigated. We find that the addition of a {\it small} 
ferromagnetic next-nearest-neighbor exchange coupling $|J_2|>0.09 J_1$ leads 
to stabilization of a 36-site unit cell VBC, although the DSL remains a 
local minimum of the variational parameter landscape. This implies that the 
VBC is not trivially connected to the DSL; instead it possesses a non-trivial 
flux pattern and large dimerization.
\end{abstract}
\pacs{75.10.Kt, 75.10.Jm, 75.40.Mg}
\maketitle

{\it Introduction}.
It is well known that, on the nonbipartite two-dimensional Kagom\'e lattice, 
the combination of low spin ($S=1/2$), low coordination number ($z=4$), and 
frustrating anti-ferromagnetic (AF) exchange interactions lead to extremely 
strong quantum fluctuations. It is, however, a widely debated and long-standing 
theoretical issue whether the ground state of the nearest-neighbor (n.n.)
spin-$1/2$ quantum Heisenberg antiferromagnet (QHAF) on the Kagom\'e lattice 
is a spin disordered state (quantum spin 
liquid),~\cite{Anderson-1973,Anderson-1987} which preserves spin rotation and 
lattice space group symmetry, or instead a valence bond crystal 
(VBC),~\cite{Marston-1991,Hastings-2000,Nikolic-2003,Singh-2007,DP-2010} which 
breaks lattice symmetries. On the experimental side, studies on a nearly 
perfect spin-$1/2$ Kagom\'e compound, Herbertsmithite 
${\rm Zn}{\rm Cu}_{3}({\rm OH})_{6}{\rm Cl}_{2}$,~\cite{Mendels-2008,Mendels-2007,Bert-2007,Nocera-2008,Nocera-2006,Shores-2007,Huang-2007,Kamenev-2008}
reveal the absence of any spin ordering down to $50$ mK despite a sizable 
n.n. AF exchange coupling ($J \approx 180$K) between spin-$1/2$ moments of 
${\rm Cu}^{2+}$. In particular, Raman spectroscopic data on Herbertsmithite 
points towards a gapless, algebraic spin liquid state.~\cite{Wulferding-2008}
This lends support to the view that the ground state of the n.n. spin-$1/2$ 
QHAF model on the Kagom\'e lattice is a long-range resonating-valence bond 
state. Within a class of variational projected wave functions, a particular 
gapless spin liquid belonging to the class of algebraic spin liquids, 
the U($1$) Dirac state, has been claimed to possess the lowest energy.~\cite{Lee-2007,Hermele-2008}
In such a state, the (mean-field) Fermi surface collapses to two 
points, where the spectrum becomes relativistic with Dirac conical excitations.
On the contrary, a recent study of the n.n. spin-$1/2$ QHAF model using 
density-matrix renormalization group (DMRG),~\cite{White-2010} establishes the
ground state to be a singlet-gapped spin liquid, supposedly with a $Z_{2}$ 
low-energy gauge structure.

From the experimental point of view, the weak ferromagnetism observed in 
Herbertsmithite has been attributed to the ferromagnetic (FM) nature of the 
next-nearest-neighbor (n.n.n.) coupling between ${\rm Cu}^{2+}$ ions in the 
Kagom\'e layers. This model was investigated in Ref.~[\onlinecite{Ma-2008}] 
by using projected wave functions and it has been found that, above a certain 
critical n.n.n. FM coupling, a gapless spin liquid with a large circular spinon
Fermi surface, named here as uniform projected Fermi sea (PFS), is stabilized 
as opposed to the $U(1)$ Dirac state. Furthermore, this state undergoes a small
dimerization, which lowers slightly its energy. The same n.n.n. FM model was 
also recently investigated using a quantum dimer model approach in 
Ref.~[\onlinecite{Ralko-2010}], showing consistently that a 36-site cell 
VBC order is favored. 

In this Rapid Communication, we revisit the spin-$1/2$ QHAF with the inclusion of n.n.n. 
exchange interactions using a more extended variational space of projected 
wave functions that may be optimized by using the technique of 
Ref.~[\onlinecite{Sorella-2006}]. In the following, we will limit to 
non-magnetic variational states and, therefore, we will not consider possible 
instabilities toward magnetically ordered states.~\cite{Lhuillier-1997} Our 
main result is that the addition of ferromagnetic n.n.n. exchange coupling 
leads to the stabilization of a 36-site unit cell VBC (in agreement with the 
results of Ref.~[\onlinecite{Ralko-2010}]) over an extended ferromagnetic 
region which starts from a very weak coupling. Although being a dimerization of
the uniform PFS, our VBC does not arise from a local instability of the latter. 
In other words, it is not trivially connected to it in the variational
parameter landscape; instead it possesses a nontrivial flux pattern and 
dimerization. Moreover, we find that the level crossing between the PFS and 
the $U(1)$ Dirac state (once suitably extended with n.n.n. hopping) occurs at 
nearly twice the value previously reported.~\cite{Ma-2008} For AF n.n.n. 
exchange coupling, the inclusion of n.n.n. hoppings in the $U(1)$ Dirac state 
leads to a considerable lowering in energy, which becomes more pronounced with 
increasing the AF coupling. Moreover, no VBC order is found in the AF n.n.n. 
coupling region.

{\it Model and wave function}.
The Hamiltonian for spin-$1/2$ quantum Heisenberg $J_{1}{-}J_{2}$ model is
\begin{equation}
\label{eqn:heis-ham}
\hat{{\cal H}} = J_{1} \sum_{\langle ij \rangle} 
{\bf \hat{S}}_{i} \cdot {\bf \hat{S}}_{j} + 
J_{2} \sum_{\langle\langle ij \rangle\rangle} 
{\bf \hat{S}}_{i} \cdot {\bf \hat{S}}_{j}
\end{equation}
where $\langle ij \rangle$ and $\langle\langle ij \rangle\rangle$ denote sums
over n.n. and n.n.n. neighbor sites, respectively. In the following, we will
consider $J_{1}>0$ and both FM and AF superexchange $J_{2}$; all energies 
will be given in units of $J_{1}$. 

The variational wave functions are defined by projecting noncorrelated 
fermionic states:
\begin{equation}
|\Psi_{{\rm VMC}}(\chi_{ij},\Delta_{ij},\mu)\rangle=
{\cal P}_{G}|\Psi_{{\rm MF}}(\chi_{ij},\Delta_{ij},\mu)\rangle,
\end{equation}
where ${\cal P}_{G}=\prod_{i}(1-n_{i,\uparrow}n_{i,\downarrow})$ is the full
Gutzwiller projector enforcing the one fermion per site constraint.
Here, $|\Psi_{{\rm MF}}(\chi_{ij},\Delta_{ij},\mu)\rangle$ is the ground state 
of mean-field Hamiltonian containing chemical potential, hopping, and pairing
terms:
\begin{equation}
\label{eqn:MF0}
{\cal H}_{{\rm MF}} =
\sum_{i,j,\alpha}(-\chi_{ij}+\mu\delta_{ij})c_{i,\alpha}^{\dagger}c_{j,\alpha}+
\Delta_{ij}c^{\dagger}_{i,\alpha}c^{\dagger}_{j,-\alpha}+{\rm h.c.}
\end{equation}
In this work, all states that we consider include hopping terms only, i.e., 
$\chi_{ij}$ up to second neighbors. The effect of including BCS pairing terms 
is discussed at the end of this Rapid Communication. Cases in which the translational symmetry
is explicitly broken will also be considered, so as to include VBC states.

Different spin-liquid and VBC phases correspond to different patterns of 
distribution of $\chi_{ij}$ and $\Delta_{ij}$ on the lattice bonds; they are
the {\it ansatz} of a given state and serve as the variational parameters that 
are optimized within the variational Monte Carlo scheme to find the 
energetically best state.~\cite{Sorella-2006} It is worth mentioning that this
method allows us to obtain an extremely accurate determination of variational 
parameters. All parameters belonging to one class (i.e., with the same 
magnitude) are generically labelled as $\chi_{\lambda}$.

{\it Results}.
We have performed our variational calculations on a 576-site (i.e., 
$36\times4\times4$) cluster with mixed periodic-antiperiodic boundary 
conditions. Such a cluster accommodates all possible VBC supercells proposed 
in the literature. In addition, it ensures non-degenerate wave functions at 
half-filling.

\begin{figure}
\includegraphics[width=0.9\columnwidth]{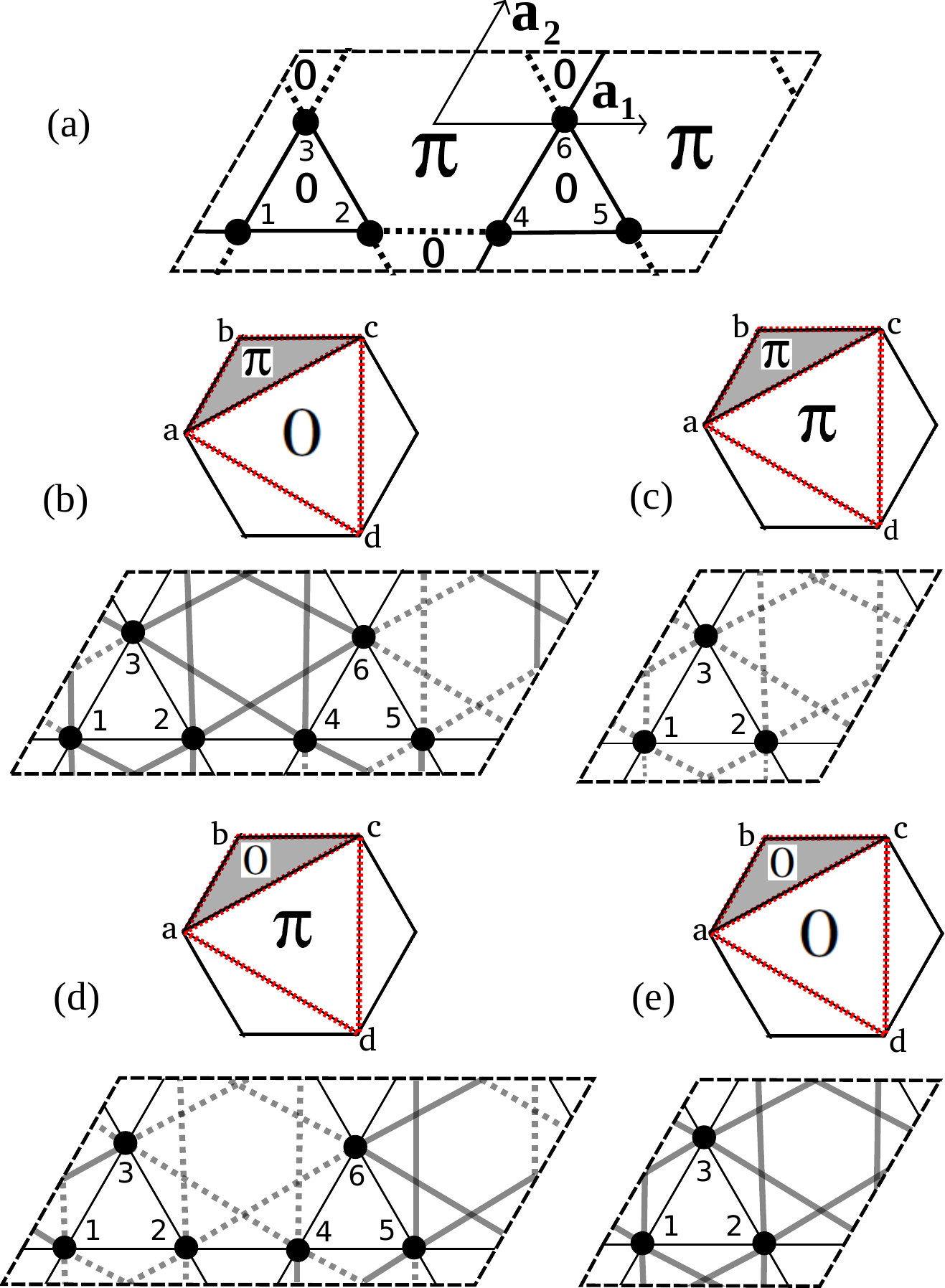}
\caption{The $U(1)$ DSL {\it ansatz}, solid (dashed) bonds denote positive 
(negative) hoppings (a). The unit cell is doubled to accommodate $[0,\pi]$ 
flux. Cases with n.n.n. hopping are also reported: the only possible 
(non-chiral) space group symmetric states built upon $[0,\pi]$ state are 
$[0,\pi;\pi,0]$ (b) or $[0,\pi;0,\pi]$ (d) and those upon the uniform $[0,0]$ 
state (i.e., PFS) are $[0,0;\pi,\pi]$ (c) or $[0,0;0,0]$ (e).}
\label{fig:pic1}
\end{figure}

For n.n. spin-$1/2$ QHAF, among the class of n.n. translationally symmetric, 
non-chiral, gapless spin \hbox{liquids}, the $U(1)$ Dirac state is given by 
the {\it ansatz} in Fig.~\ref{fig:pic1}(a). Due to flux $\varphi$ being $0$ 
and $\pi$ ($\exp{(i\varphi)}=\prod_{{\rm plaquette}}\chi_{\lambda}$) through 
the triangles and hexagons respectively, it is denoted as $[0,\pi]$. 
Its energy per site is $E/J_{1}=-0.42866(1)$. The n.n. uniform PFS state has 
no flux through any plaquette and is therefore denoted as $[0,0]$, its energy 
per site is $E/J_{1}=-0.41197(1)$.~\cite{Lee-2007,Hermele-2008} In this work
we study only gapless states in particular those with a $U(1)$ low energy gauge
structure. However, we believe that by performing a case by case projected wave
function study of all possible (a few hundred) $Z_{2}$ spin liquids on the 
Kagom\'e lattice, one can identify variationally the state found in 
Ref.~[\onlinecite{White-2010}] using DMRG. 
%%%%%%%%%%%%%%%%%%%%%%%%%%%%%%%%%%%%%%%%%%%%%%%%%%%%%%%%%%%
%%%%%%%%%%%%%%%%%%%%%%%%%%%%%%%%%%%%%%%%%%%%%%%%%%%%%%%%%%%

\begin{figure}
\includegraphics[width=0.95\columnwidth]{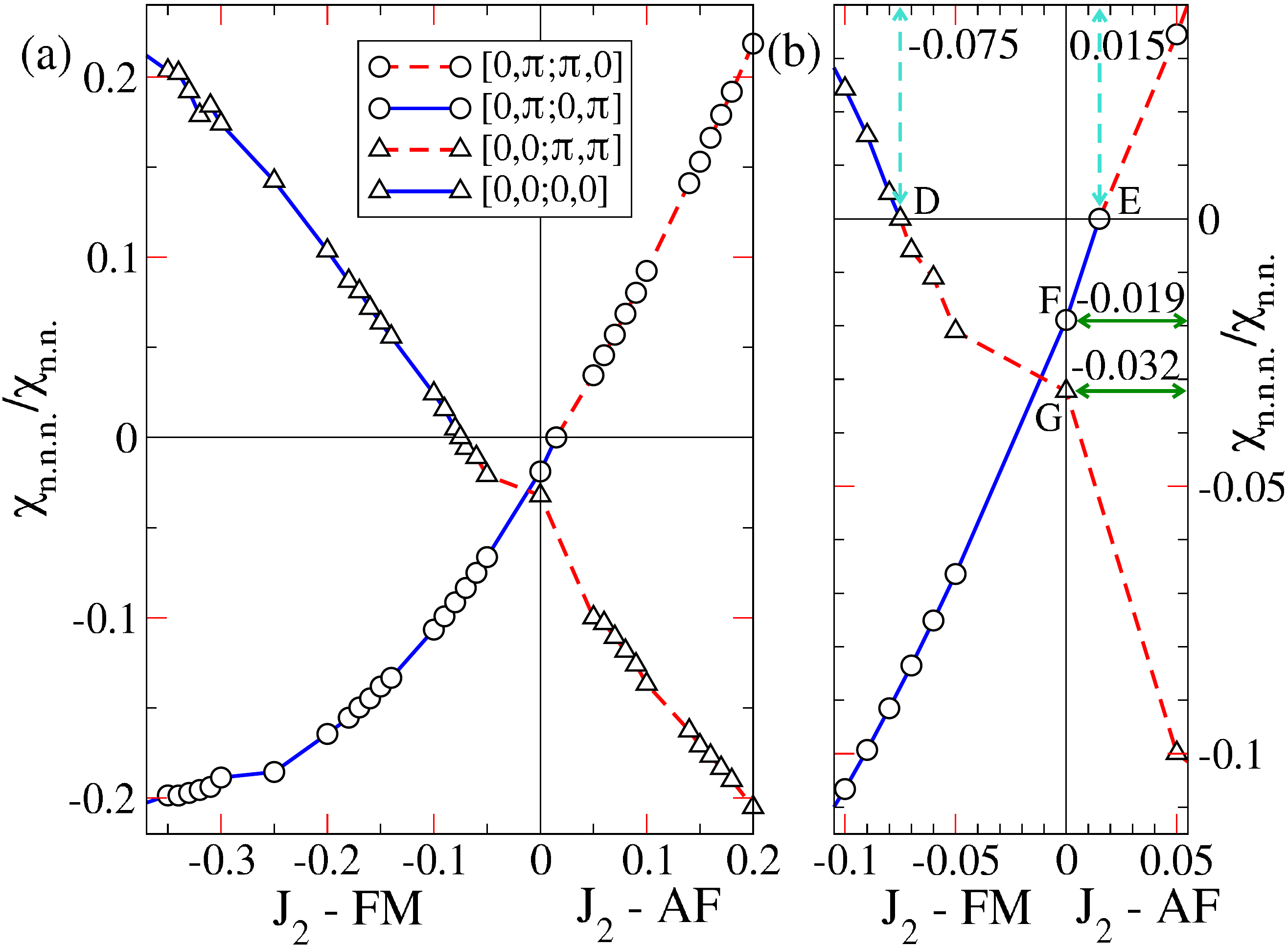}\vspace{3pt}
\includegraphics[width=0.95\columnwidth]{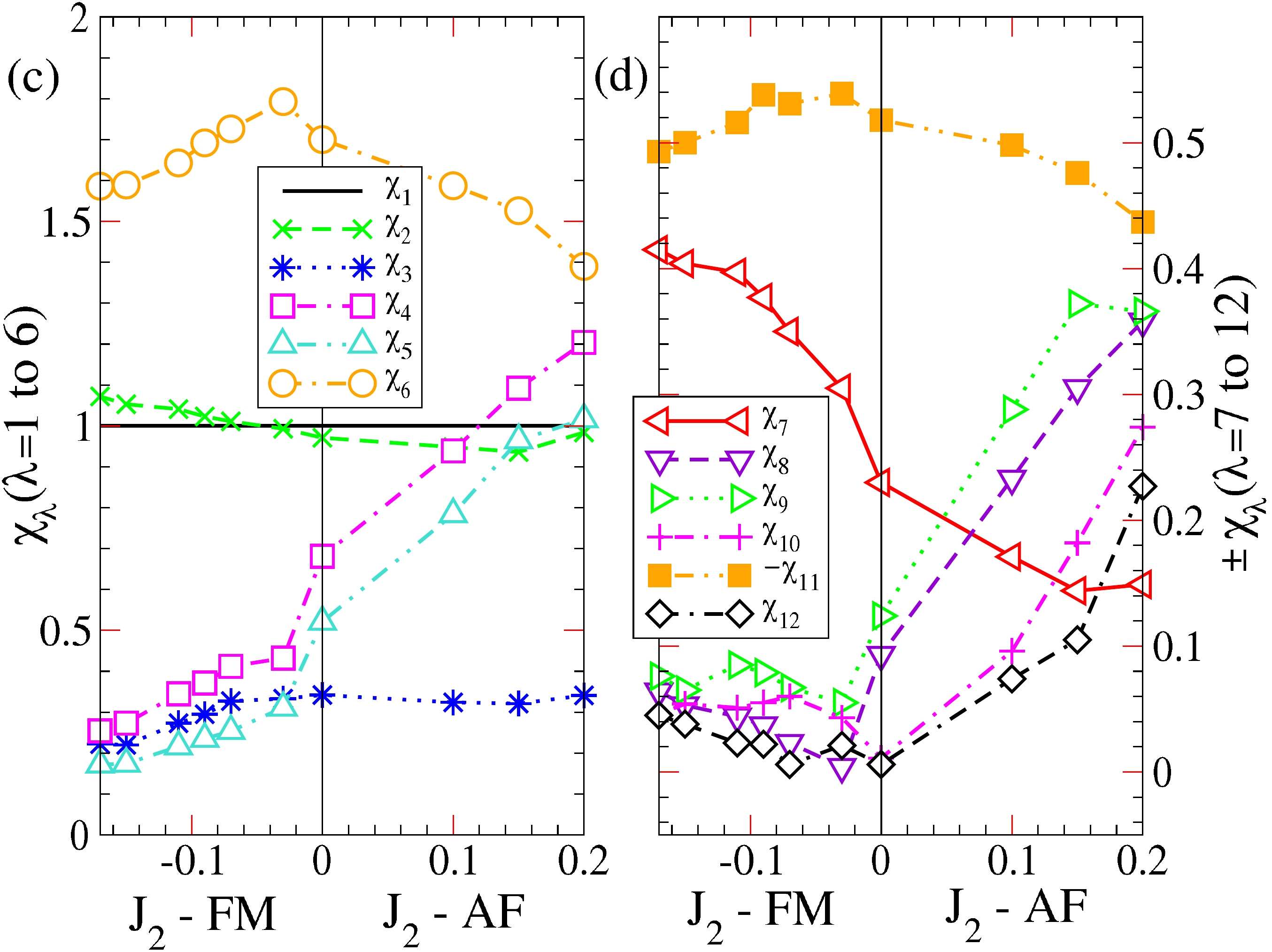}
\caption{(a) Optimized $\chi_{{\rm n.n.n.}}/\chi_{{\rm n.n.}}$ vs $J_{2}$ for 
the extended PFS and DSL states of Fig.~\ref{fig:pic1}. A zoom around $J_2=0$ 
is shown in (b). The optimized n.n. (c) and n.n.n. (d) hopping parameters
vs $J_{2}$ for the 36-site supercell VBC are also reported. $\chi_{1}=1$ is 
the reference bond. Only $\chi_{11}<0$, which implies $[0,0;\pi,\pi]$ flux in 
the P hexagons of Fig.~\ref{fig:pic4}.}
\label{fig:pic2}
\end{figure}

With the aim of investigating the effect of an additional n.n.n. exchange 
coupling (of both AF and FM type), we first extend the $[0,\pi]$ and $[0,0]$
states. While previous studies,~\cite{Ma-2008} considered wave functions 
with n.n. couplings only, here we include in addition n.n.n. couplings in the 
mean-field Hamiltonian~(\ref{eqn:MF0}), which also leads to space group 
symmetric, non-chiral, gapless spin liquids, see Fig.~\ref{fig:pic1}. 
Two new plaquettes (triangles {\it abc} and {\it acd} in Fig.~\ref{fig:pic1}) 
appear upon the inclusion of n.n.n. couplings. Space group symmetric, 
non-chiral spin liquids may now be labelled by four fluxes (but only three
are independent) (i.e., by $[\alpha,\beta;\gamma,\delta]$: $\alpha$ and $\beta$
are fluxes through original triangles and hexagons, respectively; $\gamma$ and
$\delta$ instead are fluxes through triangles {\it abc} and {\it acd}, 
respectively). The only possible states built upon the $[0,\pi]$ state are 
$[0,\pi;\pi,0]$ or $[0,\pi;0,\pi]$ and those upon the $[0,0]$ state are 
$[0,0;\pi,\pi]$ or $[0,0;0,0]$ (see Fig.~\ref{fig:pic1}). Notice that for both
DSL and PFS, the two states with different $\gamma$ and $\delta$ fluxes are
related by a change of sign in $\chi_{{\rm n.n.n.}}$. The energetically lower 
states will depend upon the actual value of the ratio $J_2/J_1$. 
This extension does not modify the topological properties associated with the 
wave functions, such as the Dirac spectrum and the large spinon Fermi surface.
Most importantly, the inclusion of n.n.n. hopping parameters leads to lowering
of the variational energies, via an optimal tuning of $\chi_{{\rm n.n.n.}}$ 
as a function of $J_{2}$ [see Fig.~\ref{fig:pic2}(a)]. It is important to note
that we purposely restrict our calculations to small enough $J_{2}/J_{1}$, since
for larger n.n.n. couplings (of both AF and FM type), it is probable that 
N\'eel states are energetically favored, and consequently our treatment
becomes insufficient.
%%%%%%%%%%%%%%%%%%%%%%%%%%%%%%%%%%%%%%%%%%%%%%%%%%%%%%%%%%%%%

\begin{figure}
\includegraphics[width=0.95\columnwidth]{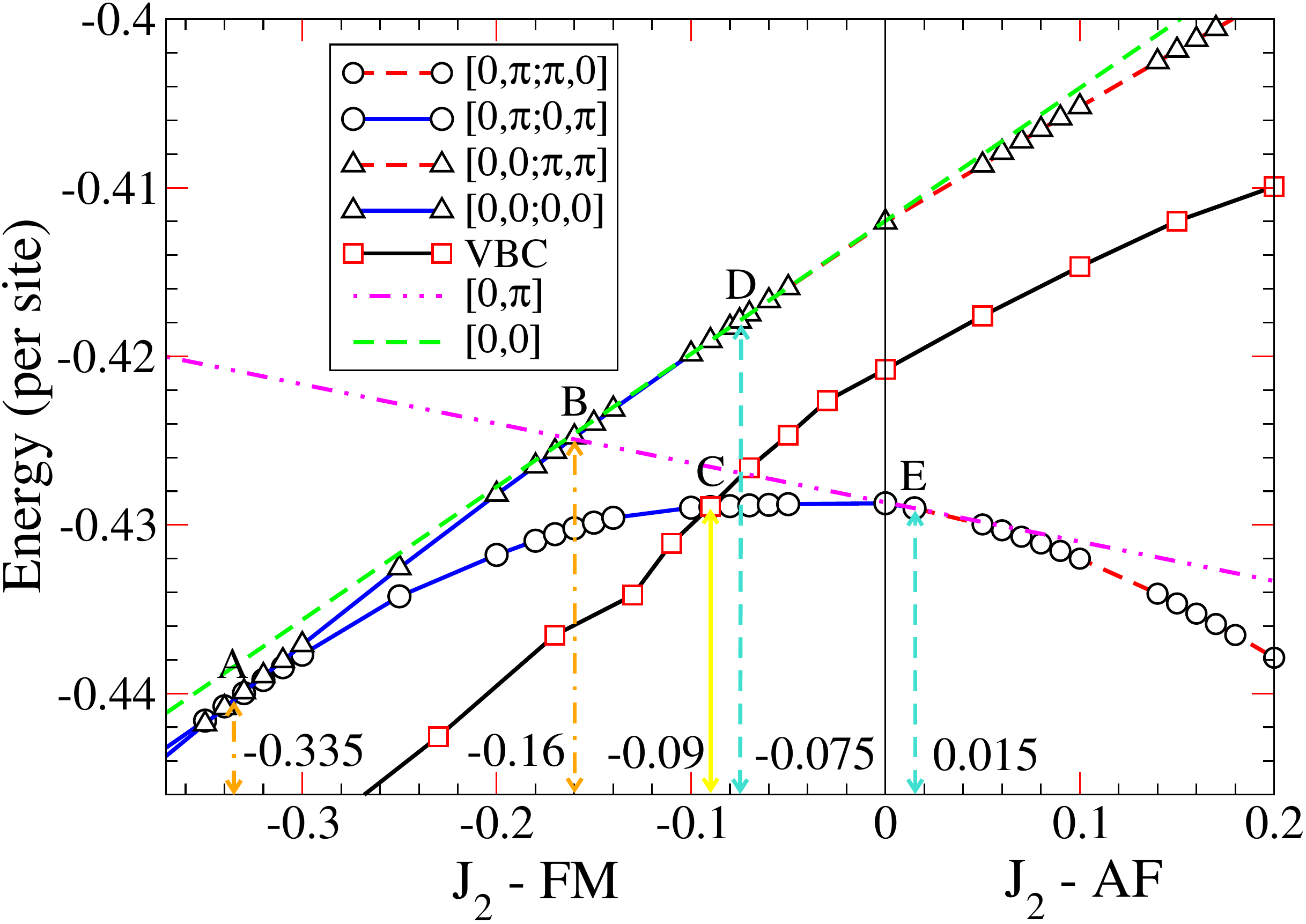}
\caption{Energy vs $J_{2}$ for spin liquid (see Fig.~\ref{fig:pic1}) and VBC 
states (see Fig.~\ref{fig:pic4}(c) and (d)).}
\label{fig:pic3}
\end{figure}
%%%%%%%%%%%%%%%%%%%%%%%%%%%%%%%%%%%%%%%%%%%%%%%%%%%%%%%%%%%%%%%
Point D in Figs.~\ref{fig:pic2}(b) and \ref{fig:pic3} marks a transition between the 
$[0,0;0,0]$ and $[0,0;\pi,\pi]$ states, and point E, the transition between 
$[0,\pi;0,\pi]$ and $[0,\pi;\pi,0]$ states, both occurring at $J_{2} \neq 0$.
Therefore, we find a finite $\chi_{{\rm n.n.n.}}$ even for the n.n. spin-$1/2$ 
QHAF [see points F and G in Fig.~\ref{fig:pic2}(b)]. We mention that these 
extended wave functions with n.n.n. hoppings lead to slightly lower energies, 
namely $E/J_{1}=-0.42872(1)$ for the $[0,\pi;0,\pi]$ state and 
$E/J_{1}=-0.41209(1)$ for the $[0,0;\pi,\pi]$ state.

Due to negative n.n.n. spin-spin correlations of $[0,\pi]$ state and positive 
for the $[0,0]$ state, a level crossing occurs at 
$J_{2}/J_{1} \approx -0.16$~\cite{Ma-2008} (see point B in Fig.~\ref{fig:pic3}). 
However, the addition of the n.n.n. hopping shifts the level crossing between 
the reference spin liquids, the $[0,0;0,0]$ and $[0,\pi;0,\pi]$ states to 
$J_{2}/J_{1} \approx -0.335$ (see point A in Fig.~\ref{fig:pic3}). 
%%%%%%%%%%%%%%%%%%%%%%%%%%%%%%%%%%%%%%%%%%%%%%%%%%
\begin{figure}
\includegraphics[width=0.9\columnwidth]{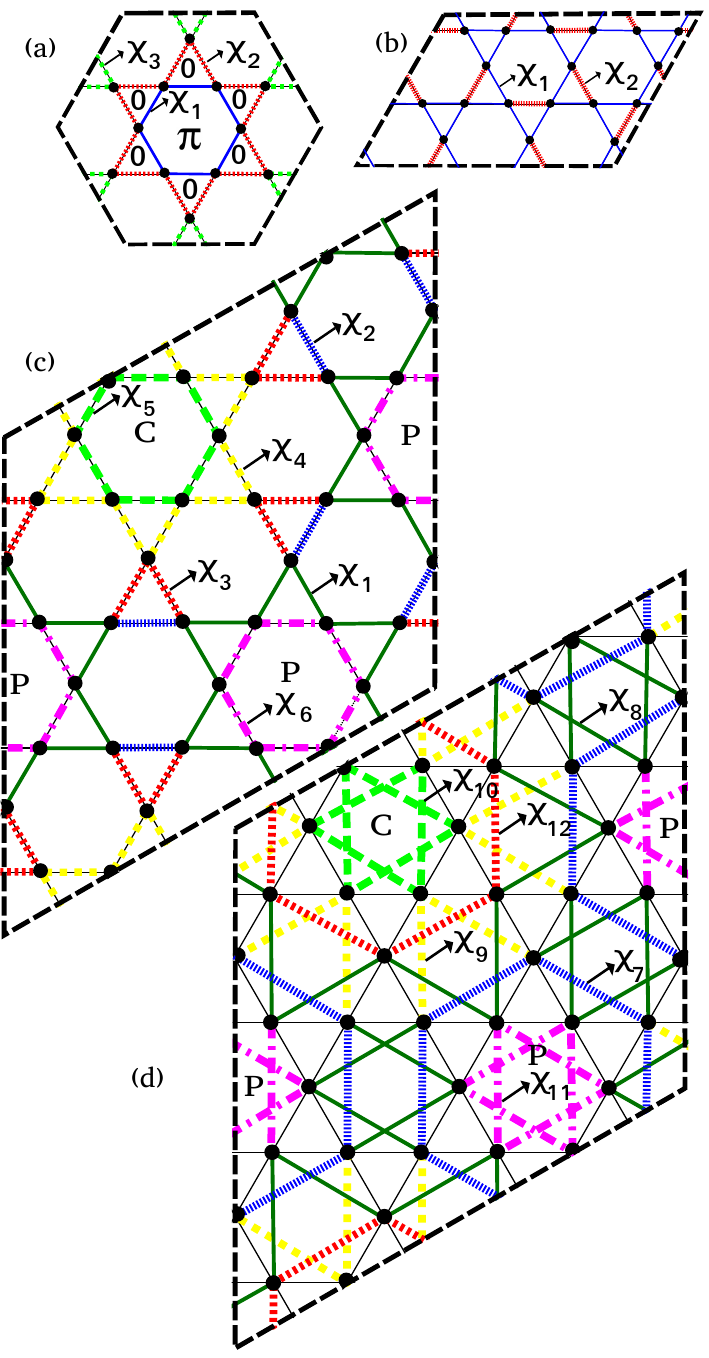}
\caption{12-site supercell (a), with three different parameters for the 
hopping; 18-site supercell, with two different parameters; 36-site supercell,
with six n.n. parameters (c), and with six n.n.n. parameters (d).}
\label{fig:pic4}
\end{figure}
%%%%%%%%%%%%%%%%%%%%%%%%%%%%%%%%%%%%%%%%%%%%%%%%55
The question of global and local instability of these spin-liquid states
towards a VBC ordering is now thoroughly addressed. In contrast to previous 
studies,~\cite{Lee-2007,Ma-2008} which aimed at checking only the local 
instabilities of spin liquids towards various dimerization patterns via the
imposition of a small bond amplitude modulation (5\%-10\%) of $\chi_{\lambda}$, 
we make a complete optimization of the parameters to detect a possible 
stabilization of VBC states. 

In the 12-site supercell, all bonds connected by $D_{6}$ operations have same 
magnitude [see Fig.~\ref{fig:pic4}(a)], leading to three classes of different 
bonds. In the 18-site supercell [see Fig.~\ref{fig:pic4}(b)], there are only
two classes of bonds. In the 36-site supercell with n.n. couplings 
[see Fig.~\ref{fig:pic4}(c)], there are six classes, given the $D_{6}$ symmetry 
about the hexagon C. By adding n.n.n. bonds and preserving this symmetry, we 
obtain six more independent bonds [see Fig.~\ref{fig:pic4}(d)]

In our analysis, we start from a large number of arbitrary different points 
(amplitude modulation of $\chi_{\lambda}$) in the variational space 
and thoroughly scan the landscape. As a consequence we find that for the n.n. 
spin-$1/2$ QHAF, the $[0,\pi]$, and $[0,0]$ states are locally and globally 
stable with respect to 12-site,~\cite{Hastings-2000} 
18-site,~\cite{Marston-1991,Ma-2008} and 
36-site~\cite{Marston-1991,Nikolic-2003,Singh-2007,DP-2010}
supercell dimerizations. Indeed, the optimization procedure always gives back
the uniform $|\chi_{\lambda}|=1$ state. We bring attention to the fact, that 
the 36-site supercell considered by us has a much larger variational space 
(six different hoppings) as compared to the ones studied in literature, which 
considered a dimerization only along the hexagon P in Fig.~\ref{fig:pic4}(c).

Upon the inclusion of a finite FM $J_{2}$, we detect the appearance of another
competing state with broken symmetries which is stabilized and is the lowest 
in energy for $J_{2}<-0.09$ (see point C in Fig.~\ref{fig:pic3}). This state 
is found to be a 36-site supercell VBC, shown in Fig.~\ref{fig:pic4}(c) and \ref{fig:pic4}(d). 
It breaks translational symmetry in the magnitude of the n.n. and n.n.n. order
parameters and the $[\gamma,\delta]$ fluxes but preserves the rotation and 
reflection symmetry in both magnitude of $\chi_{\lambda}$ and 
$[\alpha,\beta;\gamma,\delta]$ fluxes. 
The corresponding state has 12 different hopping parameters. Although we obtain
this VBC as a dimerization of the $[0,0;0,0]$ state, it is not a local 
instability of it. Instead, it possesses a large bond amplitude modulation in 
the extended variational space of n.n. and n.n.n. order parameters and selects
a flux pattern with $[\gamma,\delta]$ fluxes being $[\pi,\pi]$ in the P 
hexagons which form a honeycomb lattice, see Fig~\ref{fig:pic4}(d). On the 
contrary, all other hexagons have $[0,0]$ fluxes. The optimized 
$\chi_{\lambda}$ as function of $J_{2}$ are shown in Fig.~\ref{fig:pic2}(c) 
and \ref{fig:pic2}(d). In a previous work, which investigated the effect of a $J_{2}$ FM 
exchange coupling using projected variational wave functions,~\cite{Ma-2008} 
it was found that a dimer modulation leads to an energy minimum at 
$J_{2}/J_{1} \approx -0.16$, for approximately $4\%$ bond amplitude modulation.
In contrast, we find a different VBC wave function, which 
is stabilized starting from a very weak FM n.n.n. coupling. As mentioned above,
this state possesses a very large 36-site modulation, leading to a significant
large gain in energy.

We finish by considering the case of an AF $J_{2}$ coupling. Our study reveals
the absence of symmetry breaking, instead we find a gapless state 
with Dirac fermions, the $[0,\pi;\pi,0]$ state. Upon 
optimizing $\chi_{{\rm n.n.n.}}/\chi_{{\rm n.n.}}$, this state gets a 
significantly lower energy than the n.n. $[0,\pi]$ state, this gain becoming 
more pronounced for larger $J_{2}$ (see Fig.~\ref{fig:pic3}). Finally, the 
addition of a BCS pairing term of the $s$-wave type in the $[0,\pi;\pi,0]$ wave 
function for $J_{2}$ AF is also studied, and our calculations show that such 
an inclusion always increases the energy. However, effect of including other 
forms of pairing terms which might stabilize a gapped spin liquid or a VBC in 
the $J_{2}$ AF model is left as direction of future research. This might provide
a reconciliation with the exact diagonalization results of 
Ref.~[\onlinecite{Lhuillier-2009}], which point towards an opening of a gap 
upon addition of a small AF $J_{2}$ coupling.

In summary, we investigated the spin-$1/2$ QHAF on the Kagom\'e lattice by 
using improved variational wave functions. We found that a VBC is stabilized 
when a n.n.n. ferromagnetic superexchange coupling is considered. This state
possesses a non-trivial distribution of hopping parameters and flux pattern.


\begin{thebibliography}{16}
\expandafter\ifx\csname natexlab\endcsname\relax\def\natexlab#1{#1}\fi
\expandafter\ifx\csname bibnamefont\endcsname\relax
  \def\bibnamefont#1{#1}\fi
\expandafter\ifx\csname bibfnamefont\endcsname\relax
  \def\bibfnamefont#1{#1}\fi
\expandafter\ifx\csname citenamefont\endcsname\relax
  \def\citenamefont#1{#1}\fi
\expandafter\ifx\csname url\endcsname\relax
  \def\url#1{\texttt{#1}}\fi
\expandafter\ifx\csname urlprefix\endcsname\relax\def\urlprefix{URL }\fi
\providecommand{\bibinfo}[2]{#2}
\providecommand{\eprint}[1]{\href{}{}}

\bibitem[{\citenamefont{Anderson}(1973)}]{Anderson-1973}
\bibinfo{author}{\bibfnamefont{P. W.}~\bibnamefont{Anderson}},
  \href{http://dx.doi.org/doi:10.1016/0025-5408(73)90167-0}{\bibinfo{journal}{Mater. Res. Bull.} \textbf{\bibinfo{volume}{8}},
  \bibinfo{pages}{153 } (\bibinfo{year}{1973})}.
  

\bibitem[{\citenamefont{Anderson}(1987)}]{Anderson-1987}
\bibinfo{author}{\bibfnamefont{P.~W.}~\bibnamefont{Anderson}},
  \href{http://www.jstor.org/stable/1698247}{\bibinfo{journal}{Science.} \textbf{\bibinfo{volume}{235}},
  \bibinfo{pages}{1196 } (\bibinfo{year}{1987})}.

\bibitem[{\citenamefont{Marston and Zeng}(1991)}]{Marston-1991}
\bibinfo{author}{\bibfnamefont{J.~B.}~\bibnamefont{Marston}} and
  \bibinfo{author}{\bibfnamefont{C.}~\bibnamefont{Zeng}}, 
  \href{http://dx.doi.org/doi:10.1063/1.347830}{\bibinfo{journal}{J. Appl. Phys.} \textbf{\bibinfo{volume}{69}},
  \bibinfo{pages}{5962 } (\bibinfo{year}{1991})}.

\bibitem[{\citenamefont{Hastings}(2000)}]{Hastings-2000}
\bibinfo{author}{\bibfnamefont{M.~B.}~\bibnamefont{Hastings}}, 
  \href{http://dx.doi.org/10.1103/PhysRevB.63.014413}{\bibinfo{journal}{Phys. Rev. B.} \textbf{\bibinfo{volume}{63}},
  \bibinfo{pages}{014413 } (\bibinfo{year}{2000})}.

\bibitem[{\citenamefont{Nikolic and Senthil}(2003)}]{Nikolic-2003}
\bibinfo{author}{\bibfnamefont{P.}~\bibnamefont{Nikolic}} and
  \bibinfo{author}{\bibfnamefont{T.}~\bibnamefont{Senthil}},
  \href{http://dx.doi.org/10.1103/PhysRevB.68.214415}{\bibinfo{journal}{Phys. Rev. B.} \textbf{\bibinfo{volume}{68}},
  \bibinfo{pages}{214415} (\bibinfo{year}{2003})}.

\bibitem[{\citenamefont{Singh and Huse}(2007)}]{Singh-2007}
\bibinfo{author}{\bibfnamefont{R.~R.~P.}~\bibnamefont{Singh}} and 
  \bibinfo{author}{\bibfnamefont{D.~A.}~\bibnamefont{Huse}},
  \href{http://dx.doi.org/10.1103/PhysRevB.76.180407}{\bibinfo{journal}{Phys. Rev. B.} \textbf{\bibinfo{volume}{76}},
  \bibinfo{pages}{180407(R)} (\bibinfo{year}{2007})}.

\bibitem[{\citenamefont{Poilblanc et~al.}(2010)\citenamefont{Poilblanc, Mambrini, Schwandt}}]{DP-2010}
\bibinfo{author}{\bibfnamefont{D.}~\bibnamefont{Poilblanc}},
  \bibinfo{author}{\bibfnamefont{M.}~\bibnamefont{Mambrini}}, and
  \bibinfo{author}{\bibfnamefont{D.}~\bibnamefont{Schwandt}},
  \href{http://dx.doi.org/10.1103/PhysRevB.81.180402}{\bibinfo{journal}{Phys. Rev. B.} \textbf{\bibinfo{volume}{81}},
  \bibinfo{pages}{180402(R)} (\bibinfo{year}{2010})}; \bibinfo{author}{\bibfnamefont{D.}~\bibnamefont{Schwandt}},
  \bibinfo{author}{\bibfnamefont{M.}~\bibnamefont{Mambrini}}, and
  \bibinfo{author}{\bibfnamefont{D.}~\bibnamefont{Poilblanc}},
  \href{http://dx.doi.org/10.1103/PhysRevB.81.214413}{\bibinfo{journal}{{\it ibid}.} \textbf{\bibinfo{volume}{81}},
  \bibinfo{pages}{214413} (\bibinfo{year}{2010})}.

\bibitem[{\citenamefont{Olariu et~al.}(2008)\citenamefont{Olariu, Mendels,
  Bert, Duc, Trombe, de~Vries, and Harrison}}]{Mendels-2008}
\bibinfo{author}{\bibfnamefont{A.}~\bibnamefont{Olariu}} et al.,
  \bibinfo{author}{\bibfnamefont{P.}~\bibnamefont{Mendels}},
  \bibinfo{author}{\bibfnamefont{F.}~\bibnamefont{Bert}},
  \bibinfo{author}{\bibfnamefont{F.}~\bibnamefont{Duc}},
  \bibinfo{author}{\bibfnamefont{J.~C.} \bibnamefont{Trombe}},
   \bibinfo{author}{\bibfnamefont{M.~A.} \bibnamefont{de~Vries}},
   \bibnamefont{and} \bibinfo{author}{\bibfnamefont{A.}~\bibnamefont{Harrison}},
  \href{http://dx.doi.org/10.1103/PhysRevLett.100.087202}{\bibinfo{journal}{Phys. Rev. Lett.} \textbf{\bibinfo{volume}{100}},
  \bibinfo{pages}{087202} (\bibinfo{year}{2008})}.

\bibitem[{\citenamefont{Bert et~al.}(2007)\citenamefont{Bert, Nakamae, Ladieu,
  L'Hote, Bonville, Duc, Trombe, and Mendels}}]{Mendels-2007}
\bibinfo{author}{\bibfnamefont{F.}~\bibnamefont{Bert}} et al.,
  \bibinfo{author}{\bibfnamefont{S.}~\bibnamefont{Nakamae}},
  \bibinfo{author}{\bibfnamefont{F.}~\bibnamefont{Ladieu}},
  \bibinfo{author}{\bibfnamefont{D.}~\bibnamefont{L'H$\hat{o}$te}},
  \bibinfo{author}{\bibfnamefont{P.}~\bibnamefont{Bonville}},
  \bibinfo{author}{\bibfnamefont{F.}~\bibnamefont{Duc}},
  \bibinfo{author}{\bibfnamefont{J.-C.} \bibnamefont{Trombe}},
  \bibnamefont{and} \bibinfo{author}{\bibfnamefont{P.}~\bibnamefont{Mendels}},
  \href{http://dx.doi.org/10.1103/PhysRevB.76.132411}{\bibinfo{journal}{Phys. Rev. B} \textbf{\bibinfo{volume}{76}},
  \bibinfo{pages}{132411} (\bibinfo{year}{2007})}.

\bibitem[{\citenamefont{Mendels et~al.}(2007)\citenamefont{Mendels, Bert,
  de~Vries, Olariu, Harrison, Duc, Trombe, Lord, Amato, and
  Baines}}]{Bert-2007}
\bibinfo{author}{\bibfnamefont{P.}~\bibnamefont{Mendels}} et al.,
%  \bibinfo{author}{\bibfnamefont{F.}~\bibnamefont{Bert}},
% \bibinfo{author}{\bibfnamefont{M.}~\bibnamefont{de~Vries}},
%  \bibinfo{author}{\bibfnamefont{A.}~\bibnamefont{Olariu}},
  %\bibinfo{author}{\bibfnamefont{A.}~\bibnamefont{Harrison}},
%  \bibinfo{author}{\bibfnamefont{F.}~\bibnamefont{Duc}},
 % \bibinfo{author}{\bibfnamefont{J.}~\bibnamefont{Trombe}},
 % \bibinfo{author}{\bibfnamefont{J.}~\bibnamefont{Lord}},
 % \bibinfo{author}{\bibfnamefont{A.}~\bibnamefont{Amato}}, \bibnamefont{and}
 % \bibinfo{author}{\bibfnamefont{C.}~\bibnamefont{Baines}},
  \href{http://dx.doi.org/10.1103/PhysRevLett.98.077204}{\bibinfo{journal}{Phys. Rev. Lett.} \textbf{\bibinfo{volume}{98}},
  \bibinfo{pages}{077204} (\bibinfo{year}{2007})}.

\bibitem[{\citenamefont{Imai et~al.}(2008)\citenamefont{Imai, Nytko, Bartlett,
  Shores, and Nocera}}]{Nocera-2008}
\bibinfo{author}{\bibfnamefont{T.}~\bibnamefont{Imai}},
  \bibinfo{author}{\bibfnamefont{E.~A.}~\bibnamefont{Nytko}},
  \bibinfo{author}{\bibfnamefont{B.~M.}~\bibnamefont{Bartlett}},
  \bibinfo{author}{\bibfnamefont{M.~P.}~\bibnamefont{Shores}}, \bibnamefont{and}
  \bibinfo{author}{\bibfnamefont{D.~G.}~\bibnamefont{Nocera}},
  \href{http://dx.doi.org/10.1103/PhysRevLett.100.077203}{\bibinfo{journal}{Phys. Rev. Lett.} \textbf{\bibinfo{volume}{100}},
  \bibinfo{pages}{077203} (\bibinfo{year}{2008})}.

\bibitem[{\citenamefont{Ofer et~al.}()\citenamefont{Ofer, Keren, Nytko, Shores,
  Bartlett, Nocera, Baines, and Amato}}]{Nocera-2006}
\bibinfo{author}{\bibfnamefont{O.}~\bibnamefont{Ofer}},
  \bibinfo{author}{\bibfnamefont{A.}~\bibnamefont{Keren}},
  \bibinfo{author}{\bibfnamefont{E.~A.}~\bibnamefont{Nytko}},
  \bibinfo{author}{\bibfnamefont{M.~P.}~\bibnamefont{Shores}},
  \bibinfo{author}{\bibfnamefont{B.~M.}~\bibnamefont{Bartlett}},
  \bibinfo{author}{\bibfnamefont{D.~G.}~\bibnamefont{Nocera}},
  \bibinfo{author}{\bibfnamefont{C.}~\bibnamefont{Baines}}, \bibnamefont{and}
  \bibinfo{author}{\bibfnamefont{A.}~\bibnamefont{Amato}}, e-print
  \href{http://arxiv.org/abs/cond-mat/0610540v2}{\bibinfo{note}{arXiv:cond-mat/0610540 (2006)}}.

\bibitem[{\citenamefont{Helton et~al.}(2007)\citenamefont{Helton, Matan,
  Shores, Nytko, Bartlett, Yoshida, Takano, Suslov, Qiu, Chung
  et~al.}}]{Shores-2007}
\bibinfo{author}{\bibfnamefont{J.~S.}~\bibnamefont{Helton}} et al.,
  \bibinfo{author}{\bibfnamefont{K.}~\bibnamefont{Matan}},
  \bibinfo{author}{\bibfnamefont{M.~P.} \bibnamefont{Shores}},
  \bibinfo{author}{\bibfnamefont{E.~A.} \bibnamefont{Nytko}},
  \bibinfo{author}{\bibfnamefont{B.~M.} \bibnamefont{Bartlett}},
  \bibinfo{author}{\bibfnamefont{Y.}~\bibnamefont{Yoshida}},
  \bibinfo{author}{\bibfnamefont{Y.}~\bibnamefont{Takano}},
  \bibinfo{author}{\bibfnamefont{A.}~\bibnamefont{Suslov}},
  \bibinfo{author}{\bibfnamefont{Y.}~\bibnamefont{Qiu}},
  \bibinfo{author}{\bibfnamefont{J.-H.} \bibnamefont{Chung}},
  \bibinfo{author}{\bibfnamefont{D.~G.} \bibnamefont{Nocera}}, \bibnamefont{and}
  \bibinfo{author}{\bibfnamefont{Y.~S.}~\bibnamefont{Lee}},
  \href{http://dx.doi.org/10.1103/PhysRevLett.98.107204}{\bibinfo{journal}{Phys. Rev. Lett.}
  \textbf{\bibinfo{volume}{98}}, \bibinfo{pages}{107204}
  (\bibinfo{year}{2007})}.

\bibitem[{\citenamefont{Lee et~al.}(2007)\citenamefont{Lee, Kikuchi, Qiu, Lake,
  Huang, Habicht, and Kiefer}}]{Huang-2007}
\bibinfo{author}{\bibfnamefont{S.-H.}~\bibnamefont{Lee}} et al.,
  \bibinfo{author}{\bibfnamefont{H.}~\bibnamefont{Kikuchi}},
  \bibinfo{author}{\bibfnamefont{Y.}~\bibnamefont{Qiu}},
  \bibinfo{author}{\bibfnamefont{B.}~\bibnamefont{Lake}},
  \bibinfo{author}{\bibfnamefont{Q.}~\bibnamefont{Huang}},
  \bibinfo{author}{\bibfnamefont{K.}~\bibnamefont{Habicht}}, \bibnamefont{and}
  \bibinfo{author}{\bibfnamefont{K.}~\bibnamefont{Kiefer}},
  \href{http://dx.doi.org/doi:10.1038/nmat1986}{\bibinfo{journal}{Nature Materials} \textbf{\bibinfo{volume}{6}},
  \bibinfo{pages}{853} (\bibinfo{year}{2007})}.

\bibitem[{\citenamefont{de~Vries et~al.}(2008)\citenamefont{de~Vries, Kamenev,
  Kockelmann, Sanchez-Benitez, and Harrison}}]{Kamenev-2008}
\bibinfo{author}{\bibfnamefont{M.~A.}~\bibnamefont{de~Vries}},
  \bibinfo{author}{\bibfnamefont{K.~V.}~\bibnamefont{Kamenev}},
  \bibinfo{author}{\bibfnamefont{W.~A.}~\bibnamefont{Kockelmann}},
  \bibinfo{author}{\bibfnamefont{J.}~\bibnamefont{Sanchez-Benitez}},
  \bibnamefont{and} \bibinfo{author}{\bibfnamefont{A.}~\bibnamefont{Harrison}},
  \href{http://dx.doi.org/10.1103/PhysRevLett.100.157205}{\bibinfo{journal}{Phys. Rev. Lett.} \textbf{\bibinfo{volume}{100}},
  \bibinfo{pages}{157205} (\bibinfo{year}{2008})}.

\bibitem[{\citenamefont{Wulferding et~al.}(2010)\citenamefont{Wulferding, Lemmens,
  Scheib, R\"{o}der, Mendels, Chu, Han, Lee}}]{Wulferding-2008}
\bibinfo{author}{\bibfnamefont{D.}~\bibnamefont{Wulferding}},
  \bibinfo{author}{\bibfnamefont{P.}~\bibnamefont{Lemmens}},
  \bibinfo{author}{\bibfnamefont{P.}~\bibnamefont{Scheib}},
  \bibinfo{author}{\bibfnamefont{J.}~\bibnamefont{R\"{o}der}},
  \bibinfo{author}{\bibfnamefont{P.}~\bibnamefont{Mendels}},
  \bibinfo{author}{\bibfnamefont{S.}~\bibnamefont{Chu}},
  \bibinfo{author}{\bibfnamefont{T.}~\bibnamefont{Han}},
  \bibnamefont{and} \bibinfo{author}{\bibfnamefont{Y.~S.}~\bibnamefont{Lee}},
  \href{http://dx.doi.org/10.1103/PhysRevB.82.144412}{\bibinfo{journal}{Phys. Rev. B} \textbf{\bibinfo{volume}{82}},
  \bibinfo{pages}{144412} (\bibinfo{year}{2010})}.

\bibitem[{\citenamefont{Ran et~al.}(2007)\citenamefont{Ran, Hermele, Lee, and
  Wen}}]{Lee-2007}
\bibinfo{author}{\bibfnamefont{Y.}~\bibnamefont{Ran}},
  \bibinfo{author}{\bibfnamefont{M.}~\bibnamefont{Hermele}},
  \bibinfo{author}{\bibfnamefont{P.~A.}~\bibnamefont{Lee}}, \bibnamefont{and}
  \bibinfo{author}{\bibfnamefont{X.-G.}~\bibnamefont{Wen}},
  \href{http://dx.doi.org/10.1103/PhysRevLett.98.117205}{\bibinfo{journal}{Phys. Rev. Lett.} \textbf{\bibinfo{volume}{98}},
  \bibinfo{pages}{117205} (\bibinfo{year}{2007})}.

\bibitem[{\citenamefont{Hermele et~al.}(2008)\citenamefont{Hermele, Ran, Lee, and
  Wen}}]{Hermele-2008}
\bibinfo{author}{\bibfnamefont{M.}~\bibnamefont{Hermele}},
  \bibinfo{author}{\bibfnamefont{Y.}~\bibnamefont{Ran}},
  \bibinfo{author}{\bibfnamefont{P.~A.}~\bibnamefont{Lee}}, \bibnamefont{and}
  \bibinfo{author}{\bibfnamefont{X.-G.}~\bibnamefont{Wen}},
  \href{http://dx.doi.org/10.1103/PhysRevB.77.224413}{\bibinfo{journal}{Phys. Rev. B.} \textbf{\bibinfo{volume}{77}},
  \bibinfo{pages}{224413} (\bibinfo{year}{2008})}.

\bibitem[{\citenamefont{White et~al.}()\citenamefont{Yan, Huse, White}}]{White-2010}
\bibinfo{author}{\bibfnamefont{S.}~\bibnamefont{Yan}},
  \bibinfo{author}{\bibfnamefont{D.~A.}~\bibnamefont{Huse}}, \bibnamefont{and}
  \bibinfo{author}{\bibfnamefont{S.~R.}~\bibnamefont{White}}, e-print
  \href{http://arxiv.org/pdf/1011.6114}{\bibinfo{note}{arXiv:cond-mat/1011.6114v1 (2010)}}.

\bibitem[{\citenamefont{Ma and Marston}(2008)}]{Ma-2008}
\bibinfo{author}{\bibfnamefont{O.}~\bibnamefont{Ma}} and
  \bibinfo{author}{\bibfnamefont{J.~B.}~\bibnamefont{Marston}},
  \href{http://dx.doi.org/10.1103/PhysRevLett.101.027204}{\bibinfo{journal}{Phys. Rev. Lett.} \textbf{\bibinfo{volume}{101}},
  \bibinfo{pages}{027204} (\bibinfo{year}{2008})}.

\bibitem[{\citenamefont{Poilblanc and Ralko}(2010)}]{Ralko-2010}
\bibinfo{author}{\bibfnamefont{D.}~\bibnamefont{Poilblanc}} and
  \bibinfo{author}{\bibfnamefont{A.}~\bibnamefont{Ralko}},
  \href{http://dx.doi.org/10.1103/PhysRevB.82.174424}{\bibinfo{journal}{Phys. Rev. B.} \textbf{\bibinfo{volume}{82}},
  \bibinfo{pages}{174424} (\bibinfo{year}{2010})}.

\bibitem[{\citenamefont{Yunoki and Sorella}(2006)}]{Sorella-2006}
\bibinfo{author}{\bibfnamefont{S.}~\bibnamefont{Yunoki}} and
  \bibinfo{author}{\bibfnamefont{S.}~\bibnamefont{Sorella}},
  \href{http://dx.doi.org/10.1103/PhysRevB.74.014408}{\bibinfo{journal}{Phys. Rev. B} \textbf{\bibinfo{volume}{74}},
  \bibinfo{pages}{014408} (\bibinfo{year}{2006})}.

\bibitem[{\citenamefont{Lhuillier et~al.}(1997)\citenamefont{Lecheminant, Bernu, Lhuillier, Pierre, and
  Sindzingre}}]{Lhuillier-1997}
\bibinfo{author}{\bibfnamefont{P.}~\bibnamefont{Lecheminant}},
  \bibinfo{author}{\bibfnamefont{B.}~\bibnamefont{Bernu}},
  \bibinfo{author}{\bibfnamefont{C.}~\bibnamefont{Lhuillier}}, 
  \bibinfo{author}{\bibfnamefont{L.}~\bibnamefont{Pierre}}, \bibnamefont{and}
  \bibinfo{author}{\bibfnamefont{P.}~\bibnamefont{Sindzingre}}, 
  \href{http://dx.doi.org/10.1103/PhysRevB.56.2521}{\bibinfo{journal}{Phys. Rev. B.} \textbf{\bibinfo{volume}{56}},
  \bibinfo{pages}{2521} (\bibinfo{year}{1997})}.

\bibitem[{\citenamefont{Sindzingre and Lhuillier}(2009)}]{Lhuillier-2009}
\bibinfo{author}{\bibfnamefont{P.}~\bibnamefont{Sindzingre}} and
  \bibinfo{author}{\bibfnamefont{C.}~\bibnamefont{Lhuillier}},
  \href{http://dx.doi.org/10.1209/0295-5075/88/27009}{\bibinfo{journal}{Eur. Phys. Lett} \textbf{\bibinfo{volume}{88}},
  \bibinfo{pages}{27009} (\bibinfo{year}{2009})}.
  
\end{thebibliography}
\end{document}